\documentstyle[aps,prb,preprint,amsmath,graphicx,tighten]{revtex}
\begin{document}
\draft \title{Current-Driven Conformational Changes, Charging and
Negative Differential Resistance in Molecular Wires}
\author{Eldon G. Emberly$^1$ and George Kirczenow$^2$}
\address{$^1$NEC Research Institute, 4 Independence Way,
Princeton, NJ 08536
\linebreak
$^2$Department of Physics, Simon Fraser University, Burnaby, British
Columbia, Canada, V5A 1S6}
\date{\today}
\maketitle
\begin{abstract}

We introduce a theoretical approach based on scattering theory and
total energy methods that treats transport non-linearities,
conformational changes and charging effects in molecular wires in a unified
way. We apply this approach to molecular wires consisting of chain
molecules with different electronic and structural properties
bonded to metal contacts. We show that non-linear transport in all of these
systems can be understood in terms of a {\em single} physical mechanism
and predict that negative differential resistance
at high bias should be a generic property of such molecular wires.
\end{abstract}


\section{Introduction}

A molecular wire in its simplest definition is a single molecule that
forms an electrically conducting bridge between a pair of metallic
contacts. During the last few years molecular wires have been realized in
the laboratory and the electric current that flows through an individual
molecule in response to a potential bias applied between the contacts has
been measured\cite{Reed97,Gim99,Datta97,Gaudioso}. The current-voltage
characteristics of molecular wires are sensitive to the electronic
structure of the molecule and contacts, which depends on the spatial
arrangement (conformation) of the atoms that make up the wire and the
distribution of electronic charge along the wire: Molecular wires have
been observed to switch between conducting and non-conducting states due
to a voltage-induced redox reaction, i.e., a change of the charge
distribution in the wire\cite{Reed99}, and also in response to bias-driven
conformational changes\cite{Collier00}.  These experiments indicate that
transport non-linearities, conformational changes and charging in
molecular wires are mutually interdependent phenomena, each of which can
strongly influence the others, and that they should therefore be treated
theoretically in a unified way.  Theoretical work to date has studied some
of these phenomena independently
\cite{Datta97,DiVentra00,Ness,Yaliraki99,Sanchez00,BDTtightbinding,Emberly00,Hakkinen00}
but there has been little work directed at bringing all three of them
together within a single theoretical framework.

In this article we introduce a tractable theoretical approach that
unifies the three phenomena and present the results of self-consistent
calculations for the simplest molecular wires, namely, small chain
molecules bonded to metal electrodes. {\em At low bias} these
molecular wires are conductors or insulators depending on which atoms
make up the chain molecule and electrodes and on the number of atoms
in the chain.  Remarkably, we find that the {\em non-linear} transport
characteristics of these apparently dissimilar systems are strikingly
similar and can be understood in terms of a single physical picture:
We find that in the non-linear transport regime the molecular wire
acquires excess electronic charge. This charge resides in a molecular
orbital which, being {\em partly} occupied, lies at the Fermi level
and therefore controls electron transport through the molecule. (If
there is no molecular orbital near the Fermi level at zero bias, we
find that the conformation of the molecule changes at finite bias in
such a way that a suitable orbital appears there). As the bias
increases, the partly occupied orbital acquires more charge and the
electrostatic contribution to its energy rises together with the
electro-chemical potential of the source electrode. When this orbital
rises well above the electro-chemical potential of the drain electrode
the electron flux that it transmits {\em from drain to source}
approaches zero and ceases to decline with increasing bias while the
flux that it transmits from source to drain cannot increase further
significantly. Thus the net current through the molecular wire
saturates. At the same time a charge density wave grows along the wire
and atomic displacements occur, lengthening and thus weakening certain
molecular bonds. We find the bond weakening to be sufficient that
after saturating the current carried by the wire {\em decreases} as
the bias increases further. We predict this novel negative
differential resistance phenomenon to be a {\em generic} property of
molecular wires that involve chain molecules.

Sec.~II presents the theoretical formulation of the model that
includes charge transfer and conformational change in the non-linear
transport regime. The transport characteristics of two simple linear
chain-like molecules (homogeneous and inhomogeneous) are presented in
Sec.~III based on calculations using the model of Sec.~II. Finally,
Sec.~IV summarizes the results of this work.

\section{Theory}

Recent theories of electronic transport in molecular wires have been
based on semi-empirical tight-binding models (such as the extended
H{\"u}ckel model) and on {\em ab-initio} computations that are founded
on density functional theory. For standard values of the tight binding
model parameters adopted from the quantum chemistry literature, the
tight binding models have yielded results that are similar to (and
often in semi-quantitative agreement with) those obtained from the
{\em ab-initio} computations. For example, tight binding
calculations\cite{BDTtightbinding} have yielded current-voltage
characteristics for the benzene-dithiolate molecule bonded between
gold nano-contacts that are similar to those that have been found in
{\em ab-initio} calculations\cite{DiVentra00}. These
studies\cite{DiVentra00,BDTtightbinding} have made it evident that the
differences between the results of the tight-binding and {\em
ab-initio} calculations for this system are less significant than the
uncertainties that are due to the present limited understanding of the
conformational aspects of bonding between the molecule and metal
contacts.  Tight binding calculations of electronic transport through
molecular wires consisting of chains of carbon atoms connecting metal
contacts have also yielded results\cite{Emberly99} that that agree
remarkably well with those found in {\em ab-initio} calculations for
the same systems\cite{Lang98}.  The {\em ab-initio} calculations have
the advantage of not relying on the use of empirical
parameters. However they are much more computationally intensive than
tight-binding calculations. Therefore they are limited in their
application to smaller molecular systems and rely on the use of the
jellium approximation to treat the metallic contacts. For the same
reason it is difficult to study conformational changes using the {\em
ab-initio} techniques and the molecules were assumed to be rigid in
the above {\em ab-initio} studies.\cite{DiVentra00,Lang98} Since the
{\em ab-initio} calculations are based on density functional theory
which is known to be rigorously valid only for ground state
properties, their validity for transport calculations far from
equilibrium is uncertain {\em a priori}. Furthermore even in
calculations of ground state properties they rely in practice on the
use of the local density approximation. However, it is reasonable to
expect them to be more reliable than calculations based on simpler
mean field theories such as the Hartree-Fock approximation.  While the
Keldysh technique can in principle provide a more rigorous approach to
the problem of transport far from equilibrium, its practical
application to real molecular wires still presents major challenges at
this time.

A shortcoming of the tight-binding approach to molecular wire
transport has been that it has not provided a way to calculate the
conformational changes that occur in a molecular wire when an electric
current is driven through it, or to treat the influence of these
conformational changes and of charging of the molecule on the current
in a consistent way, within the same theoretical framework. The
formalism that we present below accomplishes this, however it should
be emphasized that it, like the {\em ab-initio} calculations mentioned
above, is based on density functional theory and should therefore also
be regarded as a practical mean field approach to the difficult
many-body problem of transport in molecular wires far from
equilibrium, and not as a rigorous treatment of this problem.

In our tight-binding formalism we represent the electronic energy
${\mathcal E}$ of the molecular wire as a functional ${\mathcal
E}[\{N_i\}]$ of the of the numbers $N_i$ of electrons occupying atomic
sites $i$ at positions $\vec{R_i}$:
\begin{eqnarray}
{\mathcal E}[\{N_i\}] = {\mathcal K}[\{N_i\}] -e\sum_{i} N_i V_{ext}(\vec{R_i})
+ \nonumber \\ \frac{1}{2}\frac{e^2}{4\pi\epsilon_0} \sum_{i\neq j} \frac{N_i
N_j}{|\vec{R_i}-\vec{R_j}|} + {\mathcal F}[\{N_i\}]
\end{eqnarray}
This is a tight-binding analog of the Kohn-Sham \cite{KohnSham}
density functional: ${\mathcal K}[\{N_i\}]$ is the tight-binding
energy of a non-interacting electron system with the same electron
distribution $\{N_i\}$ as the interacting system and $V_{ext}$ is the
corresponding external potential. The next term is the electrostatic
interaction energy between electrons on different sites. ${\mathcal
F}$ represents the exchange-correlation energy and on-site
interactions.  In the spirit of the Kohn-Sham local density
approximation \cite{KohnSham} we approximate ${\mathcal F}$ in terms
of a local function $f_i(N_i)$ that depends on the atomic species at
site $i$ as ${\mathcal F}[\{N_i\}]=\sum_{i}f_i(N_i)$. Varying
${\mathcal E}[\{N_i\}]$ with respect to $\{N_i\}$ with the
conformation $\{\vec{R_i}\}$ held fixed we obtain the tight binding
analog
\begin{equation}
\sum_{jn}t^{mn}_{ij}\psi^\lambda_{jn} + (\epsilon_{im} + \phi_{i})
\psi^\lambda_{im}  = E^\lambda \psi^\lambda_{im}
\label{TBKSEQ}
\end{equation}
of the Kohn-Sham equations for the effective one-electron eigenstates
$\Psi^\lambda = \psi^\lambda_{im}|i,m\rangle$ and energy eigenvalues
$E^\lambda$.  $t^{mn}_{ij}$ is the hopping matrix element between
orbitals $m$ and $n$ on atoms $i$ and $j$, $\epsilon_{im}$ is the bare
orbital energy,
\begin{equation}
\phi_{i} = -eV_{ext}(\vec{R_i})
+\frac{e^2}{4\pi\epsilon_0} \sum_{j\neq i}\frac{N_j}{|\vec{R_i}-\vec{R_j}|}
+\frac{\delta f_i(N_i)}{\delta N_i}
\label{TBKSPHI}
\end{equation}
$V_{ext}$ includes potentials due to the bias applied to the contacts
and image charges.  $\delta f_i(N_i)/\delta N_i$ is the variation of
the electronic site energy with the electron occupation $N_i$ of the
site.  In the numerical work reported below we approximate it by
$\delta f_i(N_i)/\delta N_i = (N_i - N^0_i)U_i$ where $N^0_i$ is the
electron occupation of the neutral atom and $U_i$ is the on site
interaction energy whose value we take from the semi-empirical
chemistry literature where it is estimated from experiment
\cite{Mcglynn72}. It is possible to also include corrections to this
term that are of higher order in $(N_i - N^0_i)$ since the relevant
empirical coefficients are available;\cite{Mcglynn72} we did not do so
in our numerical work because they are small for the species that we
considered.

For a given molecular conformation $\{\vec{R_i}\}$ and applied bias we
solve Eq.(\ref{TBKSEQ}) self-consistently with $N_i$ in
Eq.(\ref{TBKSPHI}) obtained by integrating
$\sum_{n}|\psi^\lambda_{in}|^2$ over the occupied eigenstates of
Eq.(\ref{TBKSEQ}). The electric current carried by the wire is then
given by Landauer-B\"{u}ttiker theory:
\begin{equation}
I(V) = \frac{2e}{h} \int dE\: T(E,V) [F(E,\mu_s) - F(E,\mu_d)]
\label{LB}
\end{equation}
Eq.(\ref{LB}) relates the current $I$ at finite bias voltage $V$
to the transmission probability $T(E,V)$ for an electron
to scatter from the source metallic contact through the molecule and
into the drain at an energy $E$. $F(E,\mu)$ is the Fermi function, and
$\mu_{s,d} = E_F \pm eV/2$ are the source and drain electro-chemical
potentials with $E_F$ the common Fermi energy of the contacts. We
obtain $T(E,V)$ from the eigensolutions $\Psi^\lambda$ of
Eq.(\ref{TBKSEQ}) and the velocities of electron Bloch states in the
contacts in the standard way\cite{Dattabook}.

In the Landauer picture of transport that gives rise to Eq.(\ref{LB}),
the electrons populate scattering states that flow
from deep in the contacts towards the wire according to the Fermi
functions $F(E,\mu_\alpha)$ of the contacts. These states are partly
transmitted through the wire giving rise to the current $I$ and partly
reflected. We adopt the same physical picture in calculating the {\em
actual} conformation that the molecular wire takes in the presence of
the applied bias and transmitted current: We assume that the
asymptotic regions of the contacts constitute ideal leads where the
electrons may be treated as non-interacting.  We also assume that the
filling of electron states {\em that are incident on the wire} from
each contact is described completely by the Fermi function of that
contact, i.e., it depends only on the applied bias and the temperature
and not on the conformation of the wire. [In the present work
the single particle states of Landauer theory are replaced by the
Kohn-Sham-like effective single-particle states that are the
self-consistent solutions of Eq. (\ref{TBKSEQ}).]  Then, if electron
scattering at the wire is elastic, the conformation that the wire takes is that
which minimizes the total energy of the system, the filling of the
asymptotic incident states with electrons being held fixed. The total
energy $E_{tot}(\{R_i\})$ for any trial conformation $\{R_i\}$ is
found from the self-consistent solution of Eq. (\ref{TBKSEQ}).  Its
electronic part $E_e$ is equal to ${\mathcal E}[\{\bar{N}_i\}]$ for
the conformation $\{R_i\}$ (and the fixed incident electron
population) and is given by
\begin{eqnarray}
E_e = \int E [D_s(E)F(E,\mu_s) + D_d(E) F(E,\mu_d)] dE + \nonumber \\
-\frac{e^2}{8\pi\epsilon_0}
\sum_{{i\neq j}} \frac{\bar{N}_i
\bar{N}_j}{|\vec{R_i}-\vec{R_j}|} + \sum_i \left [ -\bar{N}_i \frac{\delta
f_i(\bar{N}_i)}{ \delta \bar{N}_i} + f_i(\bar{N}_i) \right ]
\label{electronic}
\end{eqnarray}
where $\bar{N}_i$ is the self-consistent value of $N_i$ that
corresponds to the solution of (\ref{TBKSEQ}) and $D_{s,d}$ are
density of states factors for the asymptotic incident states in the
contacts. Equation (\ref{electronic}) is the non-equilibrium
tight-binding analog of Equation (2.10) of Kohn and
Sham\cite{KohnSham}: The integrand in Equation (\ref{electronic})
corresponds to the sum of the energy eigenvalues of the occupied
Kohn-Sham effective single-particle states. These single particle
states correspond to electrons that arrive at the molecular wire from
the source and drain leads and are populated according to the Fermi
distributions of the source and drain leads respectively.  The first
summation corrects for the double counting of the Coulomb energy due
to interactions between electrons that is implicit in the integral
that precedes it. The second summation corrects the contribution of
the on-site energy to the total electronic energy.

$E_e$ includes implicitly the interaction energy between
the nuclei and the electrons of the wire but omits the electrostatic
energy of the nuclei of charge $eZ_i$ interacting with each other and
with the potential $V_{ext}$ due to the applied bias and image charges in
the contacts. Thus to find the actual conformation of the wire we
minimize $E_{tot}(\{R_i\}) = E_e(\{R_i\}) +E_n(\{R_i\})$
w.r.t. $\{R_i\}$ where
\begin{eqnarray}
E_n  = \frac{e^2}{8\pi\epsilon_0}\sum_{i\neq j} \frac{Z_i
Z_j}{|\vec{R_i}-\vec{R_j}|} +   e \sum_i Z_i
V_{ext}(\vec{R_i})
\end{eqnarray}
Notice that the energy eigenvalue integral in Eq. (\ref{electronic})
is {\em independent} of the conformation of the wire. This
independence is a consequence of working with an open system with a
continuous spectrum and of our adoption of the Landauer picture of the
contacts discussed above.

\section{Predictions for some simple molecular wire systems}

\begin{figure}[!t]
\begin{center}
\includegraphics[bb = 0 100 600 700,clip,width = 0.75\textwidth]{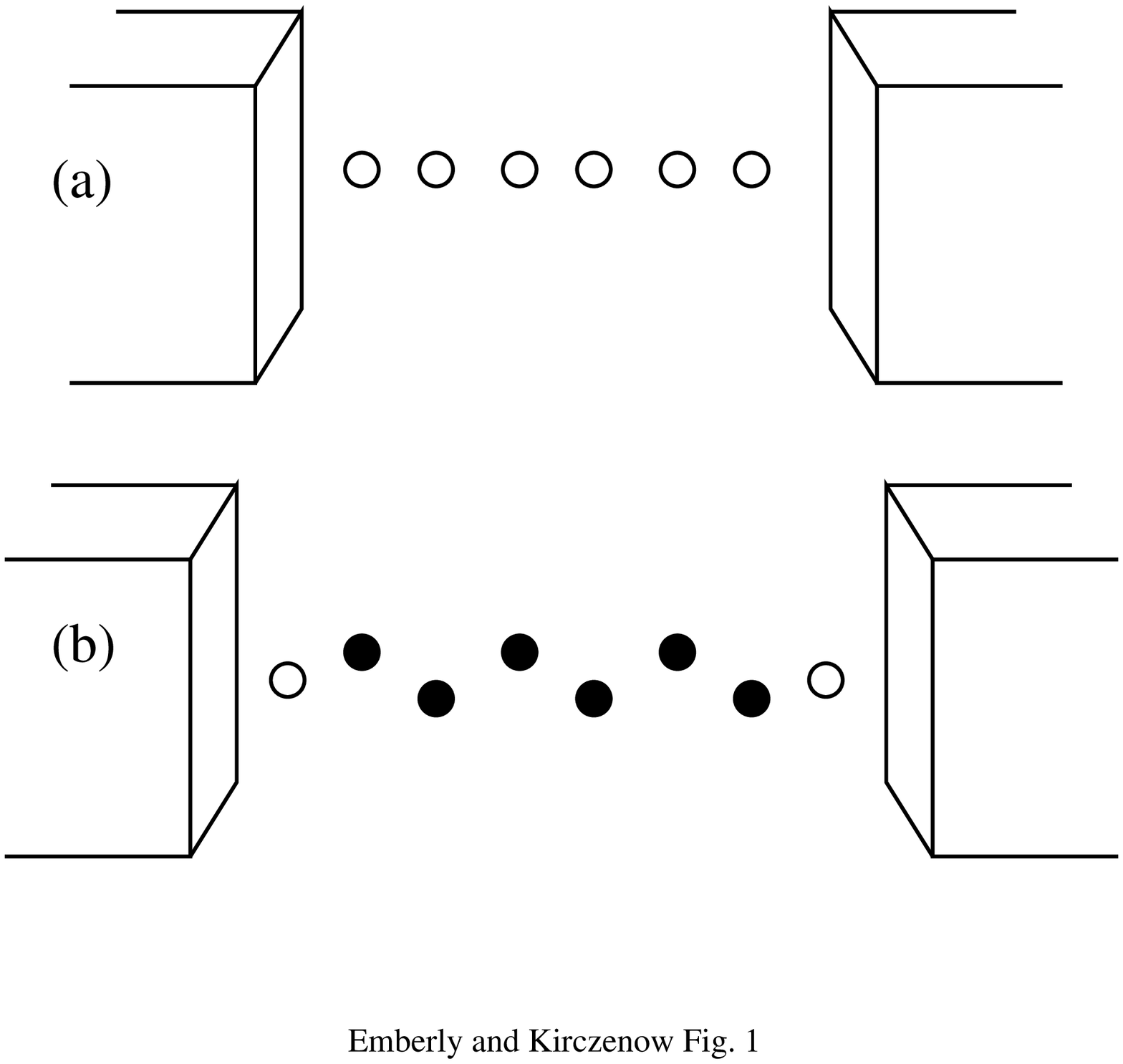}
\caption{Schematic of atomic wires bonded to ideal metallic leads. (a)
Atomic configuration of homogeneous wire. (b) Atomic configuration of
inhomogeneous wire (note the inclusion of metal atoms on the ends to
model charge transfer between the metallic contacts and the atomic
wire).}
\label{fig1}
\end{center}
\end{figure}

We have applied this methodology to two classes of simple molecular
wires consisting of chain molecules bonded to metal contacts (see
Fig.~\ref{fig1}). The contacts are modeled as single-channel ideal
leads with Fermi Energy $E_F=-10$~eV and the energy band ranging from
$-17$~eV to $-3$~eV. They are assumed to form planar equipotential
surfaces. For one class (inhomogeneous wires) the molecules are
representative of finite polyacetylene chains; we use the model
parameters of Su {\em et al.}\cite{Su79}.  Each atom of the backbone
of the chain has a single orbital with bare energy $\epsilon_C =
-11.4$~eV, nearest-neighbor hopping $t^0_C = -2.5$~eV for the
undimerized case and charging energy $U_C = -11.5$ eV. Each end of the
chain is each bonded at a perpendicular distance of $2\AA$ over a
single metal atom of a gold electrode which has a bare orbital energy
$\epsilon_{M} = -10$ eV, a charging energy $U_{M} = 8.5$ eV, and is
coupled to the end of the chain with a hopping energy $t^0_{C,M} = -2$
eV. For the other class (homogeneous wires) which is representative of
metal atomic chains bonded to metal electrodes, the unrelaxed geometry
is a linear chain with a $2.5\AA$ spacing. The same bare orbital and
charging energies are used as for the above gold metal electrode atom,
with a hopping energy $t^0_{M} = -3$ eV. As the atomic positions relax
in the course of minimizing the energy, the hopping parameters $t$
change. Following Su {\em et al.}\cite{Su79} we assume $t = t^0 +
\alpha \delta X$ where $\delta X$ is the change in bond length. For
C-C, M-C and M-M bonds, $\alpha_{C,C} = 4.125$, $\alpha_{C,M} = 2.5$
and $\alpha_{M,M} = 3.0$~eV/\AA. We calculate the current-voltage
characteristics of 6 and 7 atom inhomogeneous and homogeneous chains,
allowing the atoms to relax both in the presence and absence of a bias
voltage.

In Fig.~\ref{fig2}, the calculated site charges $-e(\bar{N}_i -
N^0_i)$ are shown for bias voltages of 0, 1 and 2V.  For the
homogeneous chains (Fig.~\ref{fig2}a,b) each atom is neutral at 0V. At
higher bias, a charge density wave (CDW) forms in both the 6 and 7
atom chains.  The CDW arises from a superposition of standing electron
waves\cite{CDW,Friedel} in the wire.  The relaxed atomic positions are the
lateral positions of the symbols in each graph.  The vertical lines
show the relaxed positions at 0V; the unrelaxed positions match the
ticks on the z axis. At 0V the 6 atom chain dimerizes while the 7 atom
chain has a frustrated dimerization pattern. At higher V, the
weakening of some bonds amplifies the standing waves and CDW's.
The inhomogeneous chains (Fig.~\ref{fig2}c,d) show similar features
although the CDW's are present even at 0V and there is a net transfer
of charge from the metal contacts to these wires. The atoms at the
ends of the chains in Fig.~\ref{fig2}c,d are metal electrode
atoms. These metal atoms are positively charged as charge is
transferred from them to the chain. For the 6 atom chain, this charge
transfer at the ends of the chain frustrates the CDW, and a charged
soliton forms. At 0 V the soliton is at the center of the chain.  It
moves away from the more negatively charged end of the chain at larger
bias.  The soliton is associated with a stretched bond. For both
inhomogeneous chains, there is considerable weakening of the bond
between the chain and drain at higher bias.  This will be seen to have
an important effect on the current flowing through the wire.

These charging and conformational changes are intimately linked
to the transport of electrons through the wires. The
voltage-dependent transmission probabilities for 0, 1 and 2V are
shown in Fig.~\ref{fig3}.  Consider first
the case of zero bias and odd atom chains (Fig.~\ref{fig3}b,d). In
isolation these chains have an odd number of valence electrons (in
this case 7) which occupy molecular orbitals,
the highest occupied molecular orbital (HOMO) being
half-filled. In an open system if this orbital were well below $E_F$
it would have to be completely filled and if it were well above the
$E_F$ it would be completely empty. Both situations lead to
unfavorable charging of the wire. Thus the HOMO must
be {\em at} $E_F$ and transmission
resonances due to the HOMO can be seen at $E_F=-10$~eV in the
two graphs. (The HOMO of the inhomogeneous wire is
more than half full due to charge transfer from
the contacts. Thus the center of the 0V HOMO resonance in
Fig.~\ref{fig3}d is slightly below $E_F$). The 6 atom
inhomogeneous chain also has a resonance near $E_F$ at zero bias
(Fig.~\ref{fig3}c). This can be explained similarly because there is
charge transfer which must go to a partly filled state near
$E_F$. This partly filled state leads to the charged soliton in
Fig.~\ref{fig2}c. However, the 6 atom homogeneous chain
has no resonance at $E_F$ at 0V
(Fig.~\ref{fig3}a) since there is no unpaired electron
and no charge transfer at zero bias. Indeed it is energetically
favorable for this chain to dimerize at 0V pushing its HOMO
lower in energy and widening the gap around $E_F$.  This explains
why there are resonances at $E_F$ at zero bias for some of the systems
but not for others.  Charge transfer also explains why there are
resonances at $E_F$ for {\it all} of the chains at non-zero bias:  All
of the wires become partly charged at non-zero bias. Thus all of them
must have a partly filled level at non-zero bias and
this level must be near $E_F$. For the 6 atom homogeneous chain, a
significant rearrangement of its atomic positions is required to
generate this state near $E_F$. (The appearance of fewer resonances than
atoms in the chain in some plots is due to degeneracy of some
of the transmitting states).

The current vs. voltage characteristics calculated from these
transmission probabilities are shown in Fig.~\ref{fig4}. Their
physical meaning is as follows: At moderate and high bias the
resonance near $E_F$ carries the net current through the wire.  As the
bias increases the wire becomes charged and the Coulomb potential due
to that charge raises the energy of the resonant state until the
bottom edge of the resonance is above the electro-chemical potential
of the drain.  Before this occurs the resonance transmits electrons
coming from {\it both} the source and drain. As the bias increases
there are fewer electrons from the drain and more from the source
going through the resonance so the current is growing. Eventually
there are no electrons from the drain going through the resonance and
the current saturates.  It then declines and negative differential
resistance sets in because the stretching and weakening of bonds
inhibits electron transmission through the wire.

\section{Summary}

In summary, we have presented theoretical work that for the first time
unifies non-linear transport, charging and conformational change in
molecular wires. We have shown that the effects of charging and
conformational change can be understood {\em together} in terms of a
remarkably robust physical mechanism of non-linear transport and
predict negative differential resistance at high bias.  One class of
molecular wires that we have modeled (the homogeneous wires) is
representative of chains of metal atoms connecting metallic contacts
such as have recently been realized in the laboratory for the case of
gold.\cite{goldchain1,goldchain2} Experiments carried out to determine
whether these systems exhibit negative differential resistance at high
bias would be of interest.

We thank R. Hill for rewarding discussions. This work
was supported by NSERC and by the Canadian Institute for
Advanced Research.


\begin{figure}[!t]
\begin{center}
\includegraphics[bb = 0 10  500 750,clip,width = 0.85\textwidth]{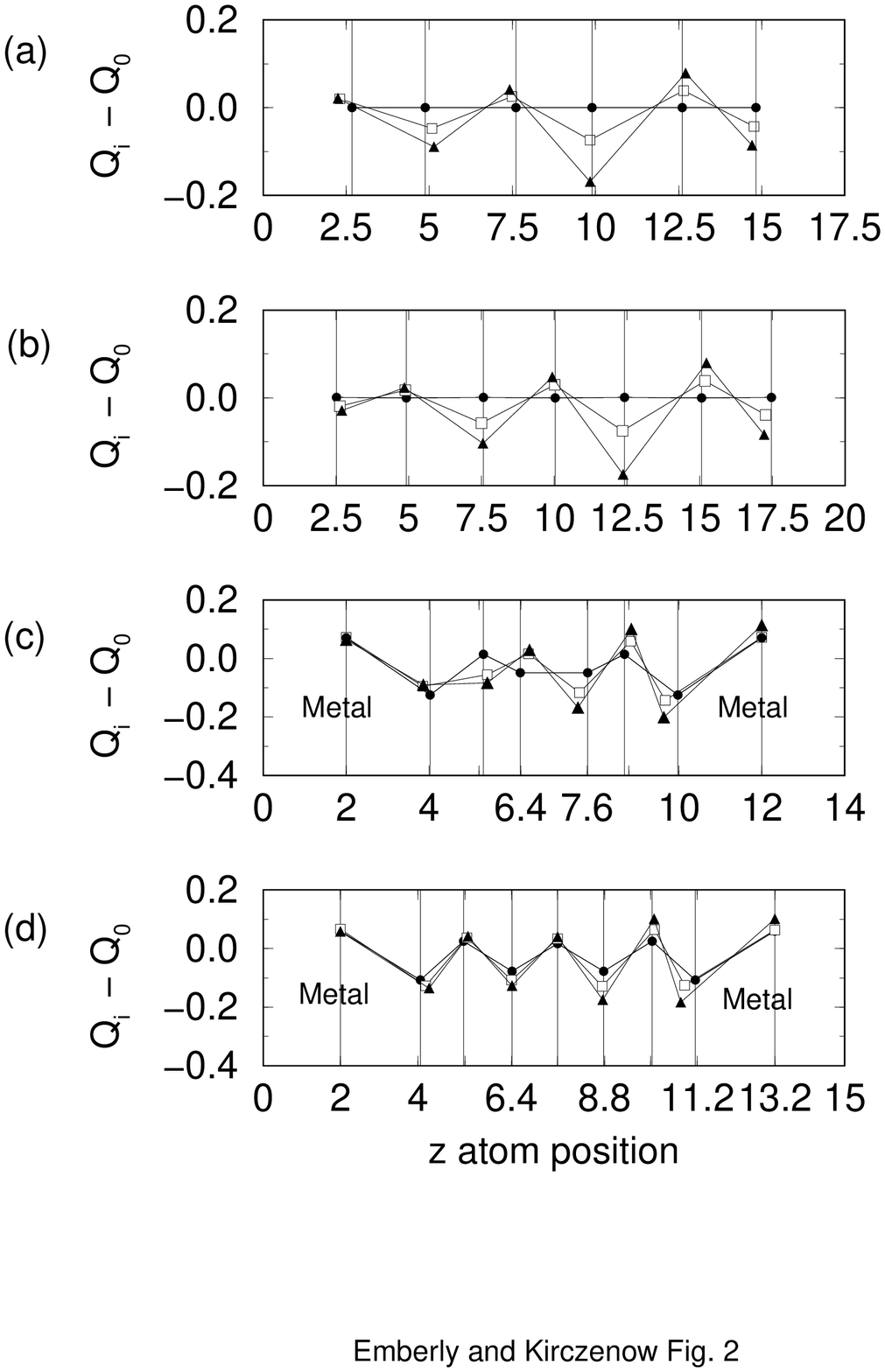}
\caption{Site charges $-e (\bar{N}_i(R_i)-N^0_i)$ in units
of $e$ and atomic positions (\AA)
for different voltages.
Circle = 0V, square = 1 V,
triangle = 2V. (a), (b): 6 and 7 atom homogeneous wire. (c), (d):
6 and 7 atom inhomogeneous wire.}
\label{fig2}
\end{center}
\end{figure}

\begin{figure}[!t]
\begin{center}
\includegraphics[bb = 0 10  500 750,clip,width = 0.85\textwidth]{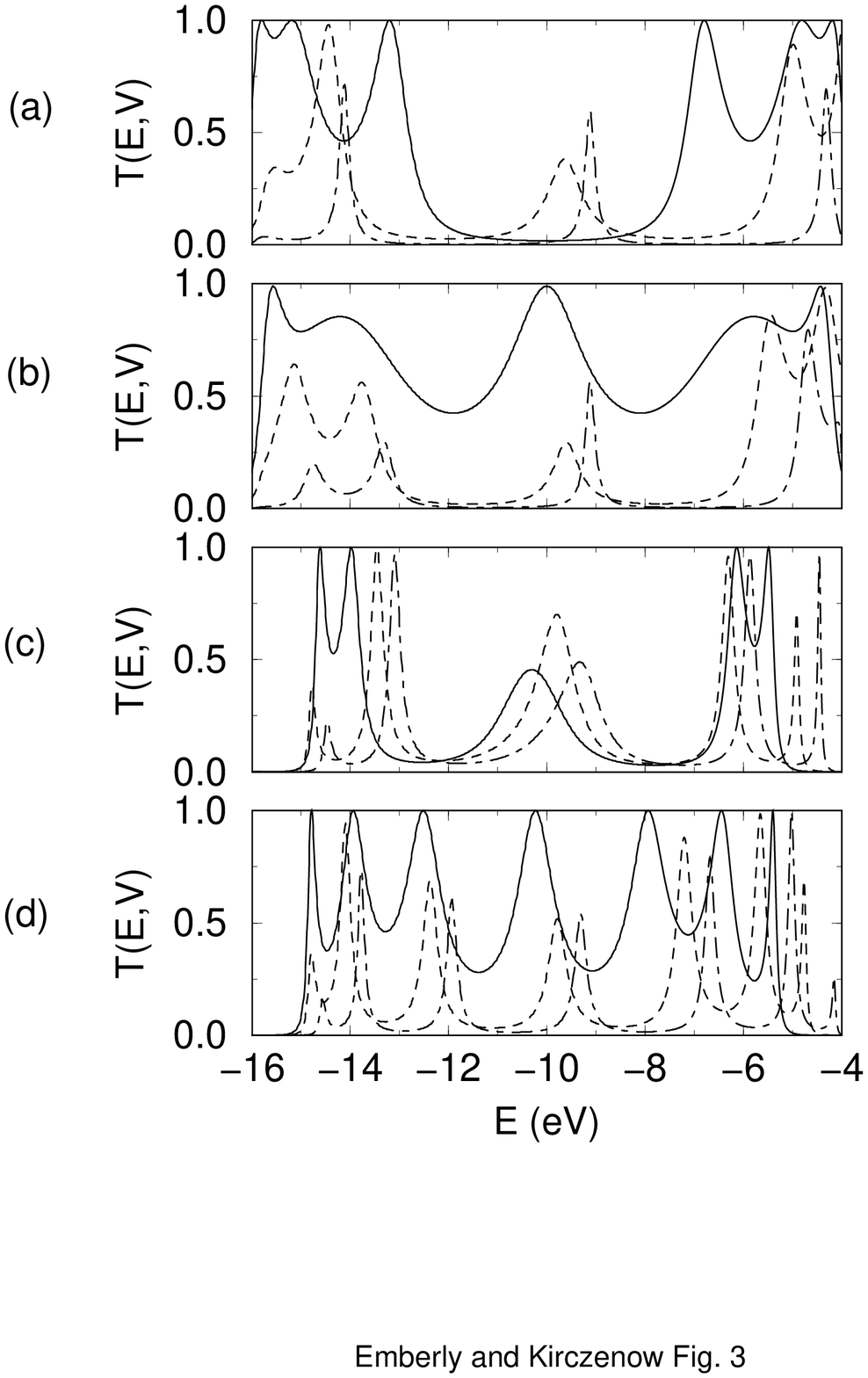}
\caption{Voltage-dependent transmission probability at 0 (solid
line), 1 (dashed) and 2V (dash-dotted). (a),(b): 6 and 7 atom
homogeneous wire. (c), (d): 6 and 7 atom inhomogeneous wire.}
\label{fig3}
\end{center}
\end{figure}

\begin{figure}[!t]
\begin{center}
\includegraphics[bb = 0 10  500 750,clip,width = 0.85\textwidth]{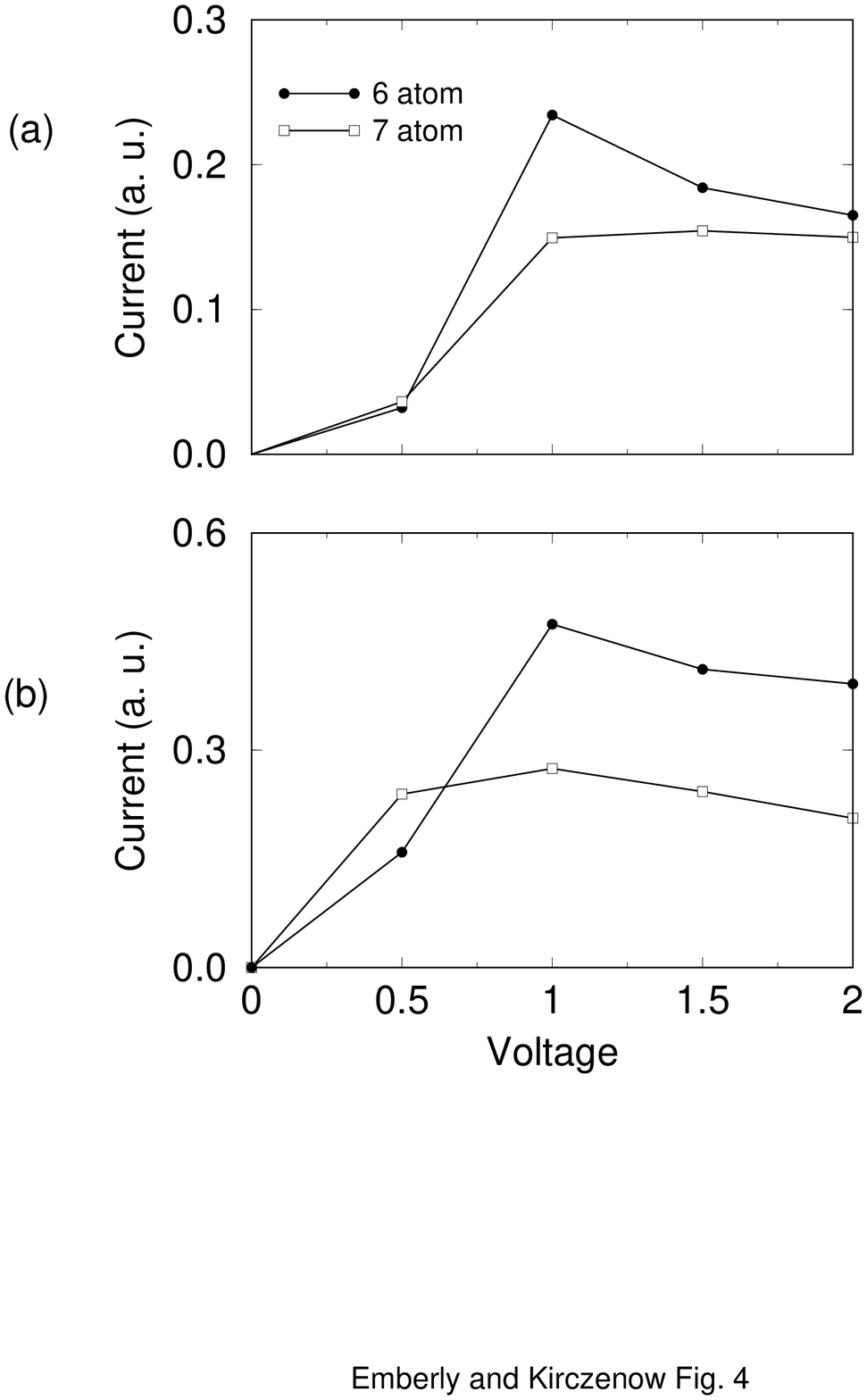}
\caption{Calculated current vs. voltage for (a) homogeneous wires and
(b) inhomogeneous wires at T = 100 K.}
\label{fig4}
\end{center}
\end{figure}



\begin{references}
\bibitem{Friedel}Notice that in Fig.~\ref{fig2}(a,b) the amplitude of the CDW
decreases as one follows the chain from its
right end (near which the electron reflections are strongest) to its left end,
as one might expect for Friedel oscillations.
\end{references}
\end{document}